\documentclass[twocolumn,prl,aps,amsmath,superscriptaddress,showpacs]{revtex4}

\usepackage{epsfig}
\usepackage{float}

\def\gb         {\beta}

\def\gd         {\delta}
\def\gee        {\epsilon}
\def\gl         {\lambda}
\def\go         {\omega}

\def\gt         {\theta}

\def\capo       {\right.\\ \left.}

\def\kk         {{\bf k}}
\def\rr         {{\bf r}}
\def\KK         {{\bf K}}
\def\ZZ         {{\bf Z}}
\def\LL         {{\bf L}}
\def\PP         {{\bf P}}
\def\TT         {{\bf \Theta }}
\def\PPi        {{\bf \Pi}}
\def\VV         {{\bf V}}
\def\WW         {{\bf W}}

\def\T         {\Theta }
\def\O         {\Omega }

\def\ck		{c_1 \kk_1}
\def\vk		{v_1 \kk_1}
\def\ckk	{c_2 \kk_2}
\def\vkk	{v_2 \kk_2}

\renewcommand{\[}{\left[}
\renewcommand{\]}{\right]}
\renewcommand{\(}{\left(}
\renewcommand{\)}{\right)}

\restylefloat{figure}

\begin{document}
%
%%%%%%%%%%%%%%%%%%%%%%%%%%%%%%%%%%%%%%%%%%%%%%%%%%%%%%%%%%%%%%%%%%%%
\title{Dynamical excitonic effects in metals and semiconductors}
\author{Andrea Marini}
\affiliation{
 Departamento de F\'\i sica de Materiales,
 Universidad del Pais Vasco, and Donostia International Physics Center.
 E--20018 San Sebasti\'an, Spain}
\author{Rodolfo Del Sole}
\affiliation{
 Istituto Nazionale per la Fisica della Materia e Dipartimento di Fisica
 dell'Universit\`a di Roma ``Tor Vergata'',
 Via della Ricerca Scientifica, I--00133 Roma, Italy}
\date{\today}

\begin{abstract} 
The dynamics of an electron--hole pair induced by the time--dependent 
screened Coulomb interaction is discussed.
In contrast to
the case where the static electron--hole interaction 
is considered we demonstrate the occurrence of important 
dynamical excitonic effects in the solution of the Bethe--Salpeter equation.
This is illustrated in the calculated absorption spectra of 
noble metals (copper and silver) and silicon.
Dynamical corrections strongly affect the spectra, 
partially canceling dynamical self--energy effects and leading to good 
agreement with experiment.
\end{abstract} 
\maketitle

The dynamics of an electron--hole pair in a many body system is a strategic 
field of research with applications to many different aspects of solid state
physics~\cite{rmp}.
For a large number of systems the observed light--absorption spectra largely
deviate from independent--particle calculations~\cite{lucia}. 
When carried out including self--energy
effects, spectra calculated for semiconductors and insulators are
characterized by a shift toward high energies with respect
to experiment.
Those deviations are corrected by including the 
electron--hole  interaction~\cite{lucia}.
The strength of these modifications 
increases as the inverse of the dielectric constant of the system.
In insulators, where the electron--hole interaction is only weakly 
screened, sharp peaks with energy below the 
optical gap (bound excitons) can be observed in the experimental 
spectra~\cite{Roessel}.
If this description is extrapolated to the metallic case
the natural conclusion is that the  electron--hole interaction has
a negligible effect on the optical spectra of metals
as the static electron--hole interaction is completely screened by the 
long--range part of the dielectric function.
This simple argument has been considered definitive  to assert
that there are no excitonic effects in metals~\cite{transient}. 
In silver and copper, however,
independent--particle calculations overestimate the experimental 
absorption spectra strengths by\,$\sim\,30\%$~\cite{silver,copper,prbnoble},
a deviation that
may be explained only in terms of many-body effects beyond the
independent--particle approximation.

The standard approach to account for the electron--hole interaction in
optical spectra is based on the solution of the 
Bethe--Salpeter equation (BSE)~\cite{rmp,strinati}\, for the two-particle
Green's function.
An important ingredient of the BSE is the  
electron--hole interaction, described by
the screened, time--dependent Coulomb interaction $W\(\rr,\rr';t-t'\)$.
However, the BSE with a time--dependent interaction is hardly
solvable~\cite{cexc}
and, as common practice,
the electron--hole interaction is assumed to be instantaneous; this is
equivalent to approximating the time Fourier transform of $W$ with its
static value, $W\(\rr_1,\rr_2,\go=0\)$.
This approximation is verified {\it a posteriori} through the comparison
with the experiment and physically 
corresponds to the assumption that the electron--hole scattering time
is much longer than the characteristic screening time of the system
(roughly speaking, the inverse of the plasma frequency).
Indeed, the static approximation is expected to
work well for transition energies much smaller than the plasma
frequency~\cite{lucia}.
However the most striking examples of systems that do
not fulfill this condition are silver and copper. 
The well--known sharp plasmon of silver, which dominates the
electron--energy--loss
spectra\,(EELS)~\cite{silver,prbnoble}, is located just
above the interband gap\,($\sim$3.9\,eV).
Similarly, the EELS 
of copper shows strong, broad peaks in the optical range~\cite{copper}.

In this letter we prove that dynamical
excitonic effects in copper and silver are possible when the 
BSE is solved with a time--dependent
electron--hole interaction.
Their experimental optical spectra
are correctly explained  only thanks to a delicate
interplay between dynamical excitonic and self--energy effects.
In order to clarify the relation between the present approach and
the previous results obtained within the static approximation~\cite{lucia}, 
we show that the optical spectrum of silicon 
can be obtained without using the approximations commonly employed
in the static approach, that is without neglecting quasiparticle
renormalization factors and dynamical
electron--hole interaction effects. 

The absorption spectrum is given by the imaginary part of the dielectric 
function
$\gee\(\go\)\equiv 1-8\pi \vec{\Lambda}^{\dagger} \PP\(\go\) \vec{\Lambda}$, 
where $\PP\(\go\)$ is the matrix 
representation of the polarization function in the non--interacting electron--hole
basis and $\vec{\Lambda}$ is a vector embodying the corresponding optical 
oscillators.
The single--particle states are calculated by means of 
density--functional--theory\,(DFT) in the 
local--density--approximation\,(LDA)~\cite{dftdetails}, while quasiparticle\,(QP)
corrections are added on top of the DFT--LDA band structure 
following the implementation of the GW method~\cite{hedin} described in Ref.~\cite{prl}.
The polarization function is obtained by solving the BSE, an integral 
equation for
the four point electron--hole Green's function $\LL\(t_1,t_2;t_3,t_4\)$.
As we are interested in the polarization function $\PP\(t\)\equiv -i 
\LL\(t,0;t,0\)$,
the BSE can be rewritten as~\cite{resonant}:
\begin{multline}
 \PP\(t\)=\PP^{\(0\)}\(t\)
  - \int dt_1\, \PP^{\(0\)}\(t-t_1\) \VV\PP\(t_1\)  \\ +
  \iint dt_1 dt_2\,
   \LL^{\(0\)}\(t,t_2;t,t_1\) \widetilde{\WW}\(t_1-t_2\)\LL\(t_1,0;t_2,0\),
\label{eq1}
\end{multline}
where the BSE kernel is  decomposed into a sum of the
instantaneous bare electron--hole and exchange interaction 
$\VV$ 
and of the time dependent screening contribution 
$\widetilde{\WW}$: $\WW\(t\)=\VV \delta (t)+\widetilde{\WW}\(t\)$.
Here we have explicitly quoted only time variables, since time dependence is
the main concern of this work.
$\LL^{\(0\)}$ is the non--interacting electron--hole Green's function
whose matrix elements are
\begin{multline}
 L_{\(cv\kk,c'v'\kk'\)}^{\(0\)}\(t_1,t_2;t_3,t_4\)=
  \gd_{v,v'}\gd_{c,c'}\gd_{\kk,\kk'}Z_{c\kk}Z_{v\kk}\\
  \gt\(t_1-t_4\)e^{-i E_{c\kk}\(t_1-t_4\)}
  \gt\(t_3-t_2\)e^{i E_{v\kk}\(t_3-t_2\)},
\label{eq2}
\end{multline}
$c$, $v$, $\kk$ being conduction, valence band and k-point indexes.
$E_{n\kk}$ (with $n=c,v$) and $Z_{n\kk}$ (smaller than 1) are the  
QP energies and renormalization factors, respectively.
The latter represent the  weights of the QP peak
in the many--body single--particle spectral function.
The more $Z_{n\kk}$ differs from 1, the more 
the high energy structures in the spectral function (like plasmonic
replicas) become important, due to
the coupling of the QP with system excitations.
Those high-energy peaks are not visible in the 
optical energy range but, nevertheless, subtract intensity
from the  QP peaks. 
Very little is known about the role played by the $Z$ factors 
in optical spectra calculations~\cite{bechstedt}.
If the $Z$ factors are included in an
independent--QP calculation~\cite{girlanda}, or even in the
BSE\,(see below),
the intensity of the resulting spectra is
strongly underestimated, both in metals and in semiconductors.
Thus the $Z$ factors are commonly set to 1 {\it by hand} in the
solution of 
the BSE or in the calculation of the independent--QP
spectra. This is, again, an approximation, needed
to reproduce the experimental results,
without a sound theoretical justification.

\begin{figure}[H]
\begin{center}
\epsfig{figure=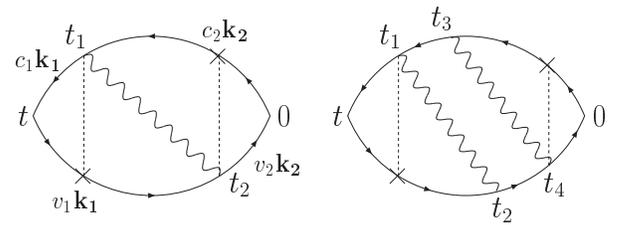,clip=,bbllx=50,bblly=650,bburx=450,bbury=800,width=8cm
}
\end{center}
\vspace{-.5cm}
\caption{
\footnotesize{
Diagrammatic representation of the first 
($\PP^{\(1\)}\(t\)$ for $t_1>t_2$, left diagram) and second 
($\PP^{\(2,a\)}\(t\)$, right diagram) order contributions to the  
polarization
function $\PP\(t\)$ according to the BSE (see text).
Crosses indicate the time points where the incoming and outgoing 
non--interacting Green's functions are ``cut" according to 
Eqs.\,(\ref{eq4a}-\ref{eq4b}). 
}}
\label{fig1}
\end{figure}

\begin{figure}[H]
\begin{center}
\epsfig{figure=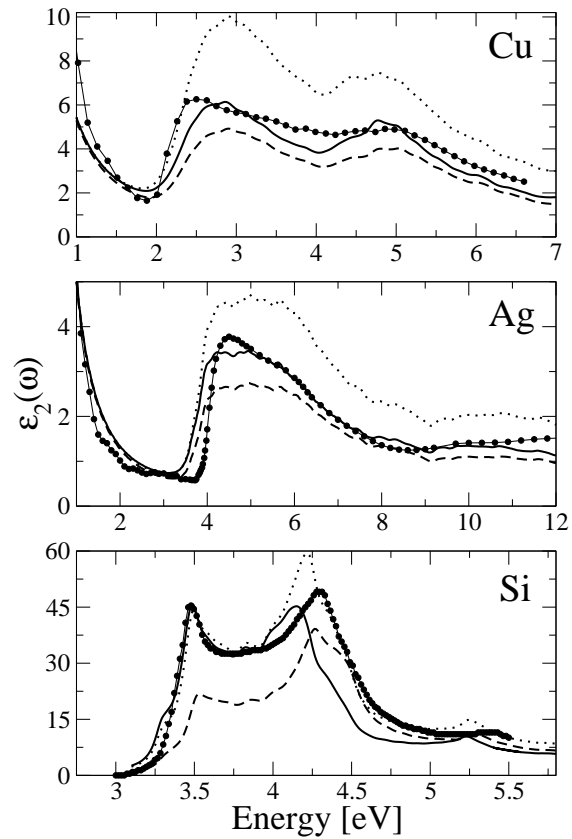,clip=,bbllx=30,bblly=30,bburx=520,bbury=765,width=7.5cm}
\end{center}
\vspace{-.5cm}
\caption{
\footnotesize{
Absorption spectrum of bulk copper, silver, and silicon.
Dotted line: SBSE without renormalization factors.
Dashed line: SBSE
including the QP renormalization factors.
Full line: result of the dynamical Bethe-Salpeter equation 
including dynamical QP and excitonic effects. Circles: 
experimental spectra (Ref.~\cite{cuexp} for Cu and Ag and
Ref.~\cite{siexp} for Si).
}}
\label{fig2}
\end{figure}

Because of the time dependent term $\widetilde{\WW}\(t_1 - t_2\)$,
Eq.\,(\ref{eq1}) cannot be rewritten in terms of $\PP\(t\)$ only.
For this reason the BSE is considered hardly solvable (if not ``practically
unsolvable"~\cite{bechstedt}) and {\it for computational convenience} 
 the electron--hole interaction is approximated as static, 
$\widetilde{\WW}\(t\)\approx\widetilde{\WW}\(\go=0\)\delta(t)$~\cite{rmp,lucia}. 
Thus 
Eq.\,(\ref{eq1}) can be formally solved by means of a Fourier transform:
\begin{align}
\PP\(\go\)=\PP^{\(0\)}\(\go\)-
 \PP^{\(0\)}\(\go\)\(\VV+\widetilde{\WW}\)\PP\(\go\).
\label{eq2a}
\end{align}
This is the static BSE\,(SBSE) commonly applied
neglecting the renormalization factors in Eq.\,(\ref{eq2}), i.e. taking
$Z_{n\kk}=1$. It yields optical spectra in good agreement with experiments
in semiconductors and insulators~\cite{lucia}. When applied to copper and
silver however, the SBSE result\,(dotted lines in Fig.\ref{fig2}) is
indistinguishable from the independent--QP calculation, 
without improving the agreement with experiment.
The inclusion of the appropriate Z factors
in the independent-QP polarization, $\PP^0$, leads to strongly
underestimated absorption spectra in all cases~\cite{refnote}. 
Hence, we look for a solution of
equation (1) without the two major approximations employed in the SBSE,
i.e. keeping the Z's smaller than 1 and $W$ frequency dependent. 
To this end we expand $\LL^{\(0\)}\widetilde{\WW}\LL$ in powers of 
$\widetilde{\WW}$. 
Using generalized indexes $\KK:=\(c\,v\,\kk\)$ the first order term of this 
expansion, $\PP^{\(1\)}\(t\)$, is given by:
\begin{multline}
P^{\(1\)}_{\KK_1\KK_2}\(t\)=\iint dt_1\,dt_2\gt\(t_1-t_2\)\\
 \[L^{\(0\)}_{\KK_1}\(t,t_2;t,t_1\)
  \widetilde{W}_{\KK_1\KK_2}\(t_1-t_2\)L^{\(0\)}_{\KK_2}\(t_1,0;t_2,0\)\capo 
+ L^{\(0\)}_{\KK_1}\(t,t_1;t,t_2\)
  \widetilde{W}_{\KK_1\KK_2}\(t_2-t_1\)L^{\(0\)}_{\KK_2}\(t_2,0;t_1,0\)\].
\label{eq3}
\end{multline}
From Eq.\,(\ref{eq2}) it is straightforward to see that
\begin{multline}
 L^{\(0\)}_{\KK_1}\(t,t_2;t,t_1\)=
 i\[P^{\(0\)}_{\KK_1}\(t-t_1\)e^{iE_{\vk}\(t_1-t_2\)}\gt\(t_1-t_2\) \capo +
 P^{\(0\)}_{\KK_1}\(t-t_2\)e^{-iE_{\ck}\(t_2-t_1\)}\gt\(t_2-t_1\)\],
\label{eq4a}
\end{multline}
\begin{multline}
 L^{\(0\)}_{\KK_2}\(t_1,0;t_2,0\)=
 i\[P^{\(0\)}_{\KK_2}\(t_2\)e^{-iE_{\ckk}\(t_1-t_2\)}\gt\(t_1-t_2\) \capo +
 P^{\(0\)}_{\KK_2}\(t_1\)e^{iE_{\vkk}\(t_2-t_1\)}\gt\(t_2-t_1\)\],
\label{eq4b}
\end{multline}
that inserted in Eq.\,(\ref{eq3}), casts $\PP^{\(1\)}\(t\)$ as a time
convolution of three terms (as shown diagrammatically in
Fig.\ref{fig1}, left diagram).
As a consequence, in the frequency domain $\PP^{\(1\)}\(\go\)$ has the
form:
\begin{align}
\PP^{\(1\)}\(\go\)=-\PP^{\(0\)}\(\go\)\[
\PPi^{\(a\)}\(\go\)+\PPi^{\(b\)}\(\go\)\]\PP^{\(0\)}\(\go\),
\label{eq6}
\end{align}
with 
$ \Pi^{\(a\)}_{\KK_1\KK_2}\(\go\)=
 \widetilde{W}^{\(+\)}_{\KK_1\KK_2}\(\go+E_{\vk}-E_{\ckk}\)$ and
$\Pi^{\(b\)}_{\KK_1\KK_2}\(\go\)=
 \widetilde{W}^{\(+\)}_{\KK_1\KK_2}\(\go+E_{\vkk}-E_{\ck}\)$,
$\widetilde{\WW}^{\(+\)}\(\go\)$ being the Laplace transform of 
$\widetilde{\WW}\(t\)$. 
The two terms denoted by $\(a\)$ and $\(b\)$ correspond to the two possible 
time orderings of the interaction ends ($t_1>t_2$ for
term (a), shown in Fig.\ref{fig1}; $t_2>t_1$ for term (b), not
shown).
Eq.\,(\ref{eq6}) can be thought of as the first order expansion 
of $\PP\(\go\)$ in the frequency--dependent interaction 
$\PPi\(\go\)=\PPi^{\(a\)}\(\go\)+\PPi^{\(b\)}\(\go\)$, which replaces 
$\widetilde{\WW}$ of the SBSE.
Thus a partial summation of the BSE can be performed writing: 
\begin{align}
\PP\(\go\)=\PP^{\(0\)}\(\go\)-
 \PP^{\(0\)}\(\go\)\[\VV+\PPi\(\go\)\]\PP\(\go\).
\label{eq8}
\end{align}
This is the Dynamical Bethe--Salpeter equation (DBSE), the central result 
of this work.
The diagrams summed up in Eq.\,(\ref{eq8}) are those containing the ladder  
series of repeated electron--hole interactions with {\it non overlapping} 
(in time) interaction lines. 
The poles of $\PP\(\go\)$, $\O_{\gl}$, will be given by the solution  of
the equation
$\[\PP^{\(0\)}\(\O_{\gl}\)\]^{-1}+\VV +\PPi\(\O_{\gl}\)=0$.
In contrast to the kernel of the SBSE, $\PPi\(\O_{\gl}\)$ is not hermitian 
and, consequently, $\O_{\gl}$ is in general complex.
Its imaginary part gives the inverse excitonic lifetime. 
Thus the interacting electron--hole states are actually dressed excitons,
or {\it quasiexcitons}.
This agrees with what has been already found in the core exciton 
limit~\cite{cexc}
and emphasizes  the analogy between the DBSE and the Dyson equation.
Consequently, as in the single--particle problem,
we expect to find similar renormalization effects on the quasiexcitonic 
Green`s function.
To develop further this aspect we expand  linearly
the smooth function $\widetilde{\WW}^{\(+\)}\(\go\)$  
around the non--interacting electron--hole energies, obtaining
$\Pi_{\KK_1\KK_2}\(\go\)\approx\Pi^{\(st\)}_{\KK_1\KK_2}+
 \T_{\KK_1\KK_2}\(\go-E_{\ckk}+E_{\vkk}\)$.
$\Pi^{\(st\)}_{\KK_1\KK_2}=
\left.\Pi_{\KK_1\KK_2}\(\go\)\right|_{\go=E_{\ckk}-E_{\vkk}}$
is the static limit of the dynamical Bethe--Salpeter kernel
which turns out to be quite similar to the kernel of the SBSE.
$\T_{\KK_1\KK_2}= \left.\partial\Pi_{\KK_1\KK_2}\(\go\)/
\partial \go\right|_{\go=E_{\ckk}-E_{\vkk}}$ are the 
 excitonic dynamical-renormalization factors.
Thus Eq.\,(\ref{eq8}) can be strongly simplified in the case of noble 
metals where  the effect of $\PPi^{\(st\)}+\VV$ is very small. 
The corresponding polarization  function
$\PP\(\go\)$ is approximatively given by:
\begin{align}
 P_{\KK_1\KK_2}\(\go\)\approx
 \frac{\[\(\ZZ^{eh}\)^{-1}+\TT\]^{-1}_{\KK_1\KK_2}}
       {\go-E_{\ckk}+E_{\vkk}+i\,0^+},
\label{eq17}
\end{align}
with $Z_{\KK_1\KK_2}^{eh}=Z_{\ck}Z_{\vk}\gd_{\KK_1\KK_2}$.
The connection between dynamical excitonic and self--energy
effects is now clear. 
$Z_{n\kk}^{-1}=1-\gb_{n\kk}$, where the negative factor
$\gb_{n\kk}$, the frequency derivative of the self-energy, is the weight
lost by the QP because
of the coupling with the excitations of $W\(\go\)$.
The excitonic factors $\TT$, instead, are due to the
modification of such coupling as a consequence of the electron--hole interaction.
Those two effects tend to cancel each other but
{\it the cancellation is, in general, not complete}, 
as exemplified  in Fig.\ref{fig2} for copper and silver.
The SBSE calculation\,(dotted line),
with $Z_{n\kk}=1$ and $\TT=0$, overestimates the
experimental intensity (circles), while the inclusion of the $Z$'s
only\,(dashed line)  underestimates it~\cite{prbnoble}.
In the DBSE (full line), obtained solving Eq.\,(\ref{eq8}), the dynamical 
$\TT$ factors 
partially compensate for the $\ZZ^{eh}$ factors yielding a 
spectral intensity in good agreement with experiment.
The case of bulk silicon is also shown in Fig.\ref{fig2}, bottom
panel.
The SBSE spectrum, calculated in the usual way by setting
the $Z$ factors to 1 {\it by hand}\,(dotted line) is in good agreement with the
experiment\,(circles). On the other hand,
the spectrum including dynamical factors in the SBSE\,(dashed line)
underestimates the experiment, as anticipated above, and the
relative intensity of the two peaks is only poorly reproduced.
The solution of the DBSE\,(solid line), instead, is in good
agreement with the experiment.
In contrast to the metallic case, however, the DBSE kernel of silicon must
contain second-order contributions in order to reproduce correctly
the experimental optical spectrum.
The main effect of the first order kernel $\PPi\(\go\)$
is indeed to balance the  reduction of optical strengths due to
self--energy
renormalization factors, as suggested by Bechstedt et
al.~\cite{bechstedt}.
However, the renormalized QP weights also imply a reduction
of the statically screened electron--hole of almost $\sim$30\%, which
is the reason for the wrong relative intensities of the two peaks
in the SBSE result. This shortcoming is fixed by the second-order
diagrams, as discussed below.
One of the two second order diagrams, $\PP^{\(2,a\)}\(t\)$,
is shown in Fig.\ref{fig1}, right diagram. In the other second-order
diagram ($\PP^{\(2,b\)}\(t\)$, not shown),
the interaction lines are ordered according to $t>t_2>t_4>t_1>t_3>0$.
Using Eqs.\,(\ref{eq4a}-\ref{eq4b}) we can isolate the external 
non--interacting polarization functions (as schematically shown in
Fig.\ref{fig1} by the crosses corresponding to the time points
$t_1$ and $t_4$).
Consequently  $\PP^{\(2,a\)}\(t\)$ can be
Fourier transformed to yield
$\PP^{\(2,a\)}\(\go\)=-
\PP^{\(0\)}\(\go\)\PPi^{\(2,a\)}\(\go\)\PP^{\(0\)}\(\go\)$
with  $\PPi^{\(2,a\)}\(\go\)=\TT^{\(a\)}\ZZ^{eh}\PPi^{\(a\)}\(\go\)$.
Together with the contribution from diagram $\PP^{\(2,b\)}$,
we obtain the total second--order kernel,  
\begin{align}
\PPi^{\(2\)}\(\go\)=
 \PPi\(\go\)+\sum_{s=a,b}\TT^{\(s\)}\ZZ^{eh}\PPi^{\(s\)}\(\go\).
\label{eq12}
\end{align}
The solution of the DBSE with the kernel given by Eq.\,(\ref{eq12})
is equivalent to summing all diagrams of the original BSE with up to two
overlapping interaction lines.
While  the first-order term ($\PPi\(\go\)$) partially restores
the optical-strength intensities, the second-order correction
($\TT^{\(s\)}\ZZ^{eh}\PPi^{\(s\)}\(\go\)$) reduces
the screening of the electron--hole pair, thus enhancing their
interaction. In the case of silicon this effect improves considerably the relative
intensity of the two main peaks, as shown in Fig.\,\ref{fig2}\,(continuous
line).
Higher-order contributions can also be included in the DBSE kernel,
although these terms contain either higher powers of $\TT$ or higher
order energy derivatives of $\PPi\(\go\)$~\cite{high} that are negligible
in the cases studied in this work.

Thus, in the case of copper and silver the BSE must be solved using a 
time--dependent electron--hole interaction to obtain good 
agreement with the experiment. The DBSE result in the case 
of silicon is similar to the case when, in the SBSE, the
QP renormalization factors are neglected~\cite{lucia}.
We understand this result as follows: the electron--hole pair, being a
neutral excitation, is less efficient than the electron and the hole alone
in exciting virtual plasmons, which is the main process leading to
QP renormalization. Only when dynamical effects are coherently
included both in the self energy and in the electron--hole interaction,
this (physically expected) result emerges from the bundle of many-body
equations. {\it This confirms the SBSE results but not the separate
approximations involved therein}.
Even though the plasma frequency of silicon is at $\sim$16\,eV 
(far above the optical energy range), the DBSE kernel needs a second order
contribution to balance the self--energy dynamical factors.

In conclusion, we have shown that,
{\it in contrast to common belief, dynamical excitonic effects in metals}
do {\it exist} and are crucial for reproducing the experimental optical
absorption.  We have demonstrated that
dynamical excitonic and self--energy effects 
must be included together in the calculation of
the response functions as they do not  completely cancel each other.
The good agreement with experiment obtained using a static
BSE kernel for semiconductors is confirmed by the present, more
general approach.

This work has been supported by the 
the EU through the NANOPHASE Research Training Network
(Contract No. HPRN-CT-2000-00167). 
We also acknowledge support from INFM PAIS CELEX and from MIUR Cofin 2002.
We thank A. Rubio and F. Bechstedt for 
helpful discussions, and C. Hogan for a critical reading.

%%%%%%%%%%%%%%%%%%%%%% REFERENCES %%%%%%%%%%%%%%%%%%%%%%%%%


\begin{thebibliography}{100}
\bibitem{refnote}
 Following Ref.~\cite{silver,prl} we include 3s/3p\,(4s/4p) core levels in
 the GW self--energy of copper\,(silver).
 Even if those states drastically modify the QP energies, they
 leave almost unchanged the $Z$ factors, and consequently, the absorption intensity.
\bibitem{high}
 We have also calculated the two third--order diagrams $\PPi^{\(3,s\)}$
 containing three overlapping interaction lines finding, respectively,
 $\PPi^{\(3,s\)}\sim\(\TT^{\(s\)}\)^2$ and
 $\PPi^{\(3,s\)}\sim\frac{\partial^2 \PPi^{\(s\)}}{\partial\go^2}$.
\end{thebibliography}
\end{document}